\documentclass[a4paper,11pt]{article}
\usepackage{jcappub} 
\usepackage{lineno}

\usepackage{dcolumn}
\usepackage{verbatim}

\usepackage{breakurl}

\usepackage{lmodern}

\pdfoutput=1
\usepackage{graphicx}
\usepackage{bm}
\usepackage{amsmath}
\usepackage{amsfonts}
\usepackage{amssymb}
\usepackage{latexsym}
\usepackage{multirow}
\usepackage{tabularx}
\usepackage{adjustbox}
\usepackage{array}
\usepackage[capitalise]{cleveref}
\usepackage{acro}
\usepackage[utf8]{inputenc}
\usepackage{CJKutf8}
\usepackage{subfigure}
\usepackage{dcolumn}
\usepackage{booktabs}
\usepackage{footnote}
\usepackage{float}
\usepackage{tikz}
\usepackage{pgfplots}
\usepackage{lineno}
\usepackage{xcolor}
\pgfplotsset{compat=newest,every axis plot/.append style={line width=1pt}}

\DeclareAcronym{LHAASO}{
  short = LHAASO ,
  long  = Large High Altitude Air Shower Observatory ,
  short-plural =  ,
}
\DeclareAcronym{GRB}{
  short = GRB ,
  long  = gamma-ray burst ,
  short-plural = s ,
}
\DeclareAcronym{CMB}{
  short = CMB ,
  long  = cosmic microwave background ,
  short-plural =  ,
}
\DeclareAcronym{EBL}{
  short = EBL ,
  long  = extragalactic background light ,
  short-plural =  ,
}
\DeclareAcronym{PDF}{
  short = PDF ,
  long  = probability distribution function ,
  short-plural = s ,
}

\DeclareAcronym{PBH}{
  short = PBH ,
  long  = primordial black hole ,
  short-plural = s ,
}

\DeclareAcronym{BH}{
  short = BH ,
  long  = Black hole ,
  short-plural = s ,
}

\DeclareAcronym{CTA}{
  short = CTAO ,
  long  = Cherenkov Telescope Array Observatory ,
  short-plural =  ,
 }
\DeclareAcronym{LHC}{
  short = LHC ,
  long  = Large Hadron Collider ,
  short-plural =  ,
 }
\DeclareAcronym{Whipple}{
  short = Whipple ,
  long  = the Whipple 10\,m telescope ,
  short-plural =  ,
 }
\DeclareAcronym{CYGNUS}{
  short = CYGNUS  ,
  long  = CYGNUS air-show array ,
  short-plural =  ,
}
\DeclareAcronym{H.E.S.S.}{
  short = H.E.S.S. ,
  long  = High Energy Stereoscopic System ,
  short-plural =  ,
}
\DeclareAcronym{Milagro}{
  short = Milagro  ,
  long  = Milagro high energy
observatory ,
  short-plural =  ,
}
\DeclareAcronym{VERITAS}{
  short = VERITAS  ,
  long  = Very Energetic Radiation Imaging Telescope Array System ,
  short-plural =  ,
}
\DeclareAcronym{Fermi-LAT}{
  short = \textit{Fermi}-LAT ,
  long  = {\it Fermi} Large Area Telescope,
  short-plural =  ,
}
\DeclareAcronym{HAWC}{
  short = HAWC ,
  long  = High Altitude Water Cherenkov Observatory  ,
  short-plural =  ,
}
\DeclareAcronym{UHE}{
  short = UHE ,
  long  = ultra high energy  ,
  short-plural =  ,
}
\DeclareAcronym{WCDA}{
  short = WCDA ,
  long  = water cherenkov detector array  ,
  short-plural =  ,
}
\DeclareAcronym{KM2A}{
  short = KM2A ,
  long  = square kilometer array  ,
  short-plural =  ,
}
\DeclareAcronym{WFCT}{
  short = WFCT ,
  long  = wide field-of-view cherenkov
telescope  ,
  short-plural =  ,
}
\DeclareAcronym{WFCTA}{
  short = WFCTA ,
  long  = wide field-of-view cherenkov
telescopes ,
  short-plural =  ,
}
\DeclareAcronym{ED}{
  short = ED ,
  long  = electromagnetic detector ,
  short-plural = s ,
}
\DeclareAcronym{MD}{
  short = MD ,
  long  = muon detector  ,
  short-plural = s ,
}
\DeclareAcronym{EAS}{
  short = EAS,
  long  = extensive air showers  ,
  short-plural =  ,
}
\DeclareAcronym{IACT}{
  short = IACT,
  long  = imaging atmospheric
cherenkov telescope  ,
  short-plural = s ,
}
\DeclareAcronym{SST}{
  short = SST,
  long  = small-sized telescope  ,
  short-plural = s ,
}
\DeclareAcronym{MST}{
  short = MST,
  long  = medium-sized telescope  ,
  short-plural = s ,
}
\DeclareAcronym{LST}{
  short = LST,
  long  = large-sized telescope  ,
  short-plural = s ,
}
\DeclareAcronym{SEM}{
  short = SEM,
  long  = standard evaporation model   ,
  short-plural =  ,
}
\DeclareAcronym{SUSY}{
  short = SUSY,
  long  = supersymmetric   ,
  short-plural =  ,
}
\DeclareAcronym{BSM}{
  short = BSM,
  long  = beyond the standard model   ,
  short-plural =  ,
}

\DeclareAcronym{SppC}{
  short = SppC,
  long  = Super proton-proton Collider  ,
  short-plural =  ,
}
\DeclareAcronym{CEPC}{
  short = CEPC,
  long  = Circular Electron Positron Collider  ,
  short-plural =  ,
}
\DeclareAcronym{ILC}{
  short = ILC,
  long  = International Linear Collider  ,
  short-plural =  ,
}
\DeclareAcronym{SWGO}{
  short = SWGO,
  long  =  Southern Wide field of view Gamma-ray Observatory,
  short-plural =  ,
}
\DeclareAcronym{FCC}{
  short = FCC,
  long  = Future Circular Collider  ,
  short-plural =  ,
}

\arxivnumber{2408.10897}
\title{Search for the Hawking radiation of primordial black holes: prospective sensitivity of LHAASO}

\author[a]{Chen Yang,}
\author[b,\ast]{Sai Wang,}
\author[a]{Meng-Lin Zhao}
\author[a,c,d,\ast]{and Xin Zhang\note[$\ast$]{Corresponding author.}}

\affiliation[a]{Key Laboratory of Cosmology and Astrophysics (Liaoning Province) \& Department of Physics, College of Sciences, Northeastern University, Shenyang 110819, China}
\affiliation[b]{Theoretical Physics Division, Institute of High Energy Physics, Chinese Academy of Sciences, Beijing 100049, China}
\affiliation[c]{National Frontiers Science Center for Industrial Intelligence and Systems Optimization, Northeastern University, Shenyang 110819, China}
\affiliation[d]{Key Laboratory of Data Analytics and Optimization for Smart Industry (Northeastern University), Ministry of Education, Shenyang 110819, China}

\emailAdd{chenyang@stumail.neu.edu.cn}
\emailAdd{wangsai@ihep.ac.cn}
\emailAdd{zhaoml@stumail.neu.edu.cn}
\emailAdd{zhangxin@mail.neu.edu.cn}

\abstract{Primordial black holes (PBHs), more generally, BHs, undergo evaporation and, in principle, will end their lives in bursts of very high-energy gamma rays. The notable aspect of the PBHs with an initial mass of $\sim10^{14}$ g is that they are expected to end their lives today. In this work, we assess the potential sensitivity of the Large High Altitude Air Shower Observatory (LHAASO) in detecting the local burst rate density of PBHs. Our results suggest that LHAASO is capable of probing for PBH bursts within a proximity of $\sim0.1$ pc from the Sun, measuring a local burst rate density of $\sim$ 1200 (or 700)$\,\mathrm{pc}^{-3}\,\mathrm{yr}^{-1}$ with $99\%$ confidence during a 3-year (or 5-year) observational campaign. This level of sensitivity surpasses the most rigorous observational constraint provided by the High Altitude Water Cherenkov Observatory (HAWC) by an order of magnitude. Additionally, we propose data analysis strategies for LHAASO to optimize the search for PBHs and reach its potential detection limits.}

\begin{document}

\maketitle
\flushbottom
\section{Introduction}

Unlike astrophysical black holes, which are remnants of stellar evolution, \acp{PBH} were produced in the early universe due to gravitational collapse of the enhanced cosmological curvature perturbations on small scales \cite{Hawking:1971ei}. {They can have a wide mass range,} spanning from the Planck mass to  several billion solar mass, contributing to various astrophysical phenomena. 
In particular, the \ac{PBH} scenario has been proposed to explain the origin of stellar-mass binary black holes reported by the LIGO and Virgo Collaborations \cite{Sasaki:2016jop,Ali-Haimoud:2017rtz}. 
Currently, the abundance of \acp{PBH} with respect to dark matter has been tightly constrained by numerous astronomical observations (for reviews see, e.g., Refs.~\cite{Carr:2020gox,Carr:2021bzv}). 
Further studies may provide key insights into both the origin of the early universe and the nature of dark matter. 

As proposed by Stephen Hawking in the 1970s \cite{Hawking:1974rv}, a black hole can radiate with a blackbody spectrum characterized by a thermal temperature inversely proportional to its mass. It emits all species of fundamental particles that have masses approximately equal to the thermal temperature \cite{MacGibbon:1990zk,MacGibbon:1991tj}.
Theoretical estimates indicate that Hawking radiation is nearly negligible for stellar-mass or more massive black holes. In contrast, it is essential for the evolution of less massive black holes, which are believed to be a smoking gun for \acp{PBH}, as these black holes can only be produced through primordial processes, not known astrophysical processes.
In particular, at the present age of the universe, a short-duration burst is expected from PBHs with an initial mass of $\sim5\times10^{14}$\,g \cite{Ukwatta:2015iba}. 
Such a burst is anticipated to emit very-high-energy gamma rays in the GeV–TeV energy range, which may be detectable by the Large High Altitude Air Shower Observatory (LHAASO) \cite{LHAASO:2019qtb}.

Once these bursts are detected, we can not only confirm the existence of \acp{PBH} and Hawking radiation, but also determine the local burst rate density of \acp{PBH}. 
If they are not detected, we {will obtain more constraining upper limits} on the local burst rate density of \acp{PBH}.
In either case, we could constrain the power spectrum of primordial density perturbations on much-smaller scales than those measured by the \ac{CMB} \cite{Wang:2019kaf}, and subsequently explore a variety of inflation models.
Furthermore, the detection of  \ac{PBH} bursts could provide insights into quantum effects of gravity and particle physics beyond the energy scale achievable by the present-day \ac{LHC} \cite{Evans:2008zzb}, as well as future facilities like the \ac{FCC} \cite{FCC:2018byv}, the \ac{CEPC} \cite{CEPCStudyGroup:2018ghi}, and the \ac{ILC} \cite{ILC:2013jhg}.

In this work, we study the prospective sensitivity of LHAASO \cite{LHAASO:2019qtb} in detection of the \ac{PBH} bursts, which are located on parsec scales from the Earth. 
Several upper bounds on the local burst rate density of \acp{PBH} have been established by various observatories, including the \ac{Whipple} \cite{Linton:2006yu}, \ac{CYGNUS} \cite{Alexandreas:1993zx}, \ac{H.E.S.S.} \cite{HESS:2023zzd}, Tibet Air Shower Array \cite{amenomori1995search}, \ac{Milagro} \cite{Abdo:2014apa}, \ac{VERITAS} \cite{Archambault:2017asc}, {\ac{Fermi-LAT}} \cite{Fermi-LAT:2018pfs}, as well as \ac{HAWC} \cite{HAWC:2019wla}. 
Compared with these existing programs, \ac{LHAASO} is sensitive to broader energy range and offers higher sensitivity \cite{LHAASO:2019qtb}. 
The detectable energy range for gamma rays at \ac{LHAASO} spans from 30\,GeV to 100\,PeV.
The lower energy threshold facilitates the possible detection of \ac{PBH} bursts at greater distances, while the higher energy limit expands the field of very high-energy gamma-ray astronomy beyond the currently observable bands.
Furthermore, \ac{LHAASO}'s sensitivity is significantly greater than that of existing programs when the gamma-ray energy exceeds 10\,TeV.
Therefore, we expect this program to explore unknown territories by either establishing the strongest upper limits on the local burst rate density of \acp{PBH} or achieving the first detection of \ac{PBH} bursts.

The remainder of this paper is organized as follows. In Section~\ref{sec:2}, we provide a brief summary of the theory of \ac{PBH} bursts and the data analysis methods employed. In Section~\ref{sec:3}, we present prospective constraints from the \ac{LHAASO} on the local burst rate density of \acp{PBH}. Finally, in Section~\ref{sec:5}, we offer conclusions and a discussion.

\section{PBH bursts: theory and data analysis}\label{sec:2}

In this section, we summarize the theory of \ac{PBH} bursts and the method of data analysis.

\subsection{Time-integrated number of photons per unit energy for an evaporating PBH}

Throughout this work, we adopt the \acl{SEM}, which posits that the particles emitted from \acp{PBH} due to Hawking radiation consist  exclusively of the particles in the Standard Model \cite{MacGibbon:1990zk,MacGibbon:1991tj}. 
The temperature $T$ of a spinless \ac{PBH} is inversely proportional to its mass $M$, specifically given by $T=M_{P}^{2}/(8\pi M)$, where $M_{P}$ denotes the Planck mass. 
As the \ac{PBH} undergoes evaporation, its mass decreases while its temperature increases, resulting in a greater variety and quantity of particle species.
During the final stage of \ac{PBH} evaporation, a significant outburst of particles occurs, leading to a burst event.
The final particle products from such a burst include photons, electrons, protons, neutrinos, and their antiparticles. 
In this work, we focus on the very-high-energy gamma rays, which are potentially detectable  by \ac{LHAASO}.

To express the time-integrated number of photons per unit energy, it is convenient to represent the temperature of \ac{PBH} in terms of the burst duration $\tau$, corresponding to the \ac{PBH}'s remaining lifetime \cite{Halzen:1991uw}:
 \begin{equation}\label{eq:ttau}
T \simeq 7.8 \times 10^3 \times \left(\frac{1 \mathrm{~s}}{\tau}\right)^{\frac{1}{3}}\,\mathrm{GeV}\ .
\end{equation}
Considering the observational energy range of \ac{LHAASO} \cite{LHAASO:2019qtb}, we focus on an energy range $T>10$\,GeV in this work, taking into account both primary and secondary particles. 
During the burst duration of a given \ac{PBH}, the time-integrated number of photons per unit energy is approximated as follows \cite{Petkov:2008rz}:
\begin{equation}\label{eq:photon spetra}
\frac{d N_{\gamma}}{d E}(E;\tau) \approx 9 \times 10^{35} \times \left\{\begin{aligned}
& \left(\frac{1 \mathrm{~GeV}}{T}\right)^{\frac{3}{2}}\left(\frac{1 \mathrm{~GeV}}{E}\right)^{\frac{3}{2}} \,\mathrm{GeV}^{-1}  \quad\quad  \text { for } E<T, \\
& \left(\frac{1 \mathrm{~GeV}}{E}\right)^3 \,\mathrm{GeV}^{-1} \quad\quad\quad\quad\quad\quad\quad~   \text { for } E \geq T,
\end{aligned} \right. 
\end{equation}
where $E$ is the energy of the photons, $T$ is related to $\tau$ via Eq.~(\ref{eq:ttau}), and $N_{\gamma}$ represents the time-integrated number of photons.

\subsection{Data analysis}

Assuming that a \ac{PBH} burst is within the field of view of a given detector, we expect the total number of photons received by the detector to be given by
\begin{equation} \label{eq:mu}
\mu(r, \theta_i, \tau)=\frac{(1-f)}{4 \pi r^2} \int_{E_1}^{E_2} \frac{d N_{\gamma}}{d E} (E;\tau)A_{\gamma}(E, \theta_i) d E\ , 
\end{equation} 
where $r$ is the distance from the detector to the burst, $\theta_i$ is the zenith angle of band $i$, $f$ is the dead-time fraction of the detector,\footnote{Throughout this work, we disregard effects of the dead-time fraction by fixing $f=0$.} and $A_{\gamma}$ is the effective area for photon detection. 
Additionally, we assume that the duration of \ac{PBH} burst is sufficiently short that all particles are detected from a zenith angle within the same band, eliminating the need to consider the Earth's rotation, which would be relevant for detecting long-lasting bursts.

To claim a detection of \ac{PBH} bursts or to place upper limits on the local burst rate density, we must establish a minimum number of photons $\mu_{\mathrm{min}}(\theta_{i},\tau)$, serving as a threshold for asserting an excess over the background at $5\sigma$ confidence level. 
First, we we define the number of trials $N_{t}$ trials as follow \cite{Lopez-Coto:2021lxh}: 
\begin{equation}
N_{\mathrm{t}}=\frac{S}{\tau}\left(\frac{\theta_{\text {fov}}}{\theta_{\text {res}}}\right)^2\ , 
\end{equation}
where $S$ is an observation duration, and $\theta_{\text{fov}}$ and $\theta_{\text{res}}$ denote the field-of-view and angular resolution of the detector, respectively.
Next, considering a correction for $N_{t}$ trials, we define a $5\sigma$ detection as the number of counts $n$ that corresponds to a Poisson probability $P$ associated with a post-trial p-value given by
\begin{equation}\label{eq:posttrial}
p_{c}=\frac{p_{0}}{N_{\mathrm{t}}}=P(\geq n | n_{\mathrm{bk}}) \ ,
\end{equation}
where $p_{0}=2.3\times10^{-7}$ is the pre-trial p-value corresponding to $5\sigma$.
Here, $P(\geq k | m)$ is the Poisson probability of obtaining at least $k$ counts when the Poisson mean is $m$. 
The number of background counts $n_{\mathrm{bk}}$ is the product of the burst duration $\tau$ and the background rate $R_{b}(\theta_{i})$, expressed as 
\begin{equation}
n_{\mathrm{bk}}=\tau R_{b}(\theta_{i})\ .
\end{equation}  
Here, the background rate is given by \cite{Abdo:2014apa}
\begin{equation} \label{eq:R}
R_b(\theta_i)=\int_{E_1}^{E_2} \frac{d N_p}{d E}(E) A_p(E, \theta_i)d E \times\left[2\pi(1-\cos\theta_{\mathrm{res}})\right]\times 1.2\ , 
\end{equation}
where $A_{p}$ is the effective area for protons, the cosmic ray (proton) spectrum is given by \cite{Panov:2006kf} 
\begin{equation}
\frac{d N_p}{d E}(E) =7900 \times \left(\frac{E}{1\,\mathrm{GeV}}\right)^{-2.65}\mathrm{m}^{-2} \mathrm{s}^{-1}\mathrm{sr}^{-1}\mathrm{GeV}^{-1}\ .
\end{equation} 
The correction factor of 1.2 accounts for the influence of other cosmic ray particles on the background. 
Finally, $\mu_{\mathrm{min}}(\theta_{i},\tau)$ is associated with a 50\% Poisson probability of detecting a $5\sigma$ excess, implying that we can determine  $\mu_{\mathrm{min}}(\theta_{i},\tau)$ by solving 
\begin{equation}
P(\geq n | n_{\mathrm{bk}}+\mu_{\mathrm{min}}(\theta_{i},\tau))=0.5\ ,
\end{equation} 
indicating that the Poisson probability $P$ of obtaining at least $n$ counts is 50\%.

By setting the left hand side of Eq.~(\ref{eq:mu}) equal to $\mu_{\min}(\theta_i, \tau)$, we can determine the maximum distance from which a \ac{PBH} burst can be detected by the detector: 
\begin{equation}\label{eq:radius}
r_{\max }\left(\theta_i, \tau\right)=\left[{\frac{(1-f)}{4 \pi \mu_{\min}\left(\theta_i, \tau\right)} \int_{E_1}^{E_2} \frac{d N_{\gamma}}{d E}(E;\tau) A_{\gamma}(E, \theta_i) d E}\right]^{\frac{1}{2}}\ .
\end{equation}
Consequently, an effective detectable volume for the detector is given by 
\begin{equation}\label{eq:volume}
V(\tau)=\sum_i V\left(\theta_i, \tau\right)=\frac{4}{3} \pi \sum_i r_{\max }^3\left(\theta_i, \tau\right) \frac{\mathrm{FOV}\left(\theta_i\right)}{4 \pi}\ ,
\end{equation} 
where the solid angle corresponding to the field of view associated with the band ${i}$ is defined as
\begin{equation}\label{eq:fov}
\mathrm{FOV}\left(\theta_i\right)=2 \pi\left(\cos \theta_{i, \min }-\cos \theta_{i, \max }\right) \ .
\end{equation}
Here, $\theta_{i, \min }$ and $\theta_{i, \max }$ represent the minimum and maximum zenith angles of band $i$, respectively.

To establish an upper limit on the local burst rate density, we assume that \acp{PBH} are uniformly distributed in the vicinity of our Sun. 
The Poisson probability for a null detection is $P(0 | m)=1-P(n \geq 1 | m)=1-X={m^0 e^{-m}}/{0 !}$, where $m$ represents the upper limit on the expected number of \ac{PBH} bursts at a confidence level $X$. 
For $X = 99\%$, we find $m = \ln 100 \approx 4.6$. 
This result leads to the 99\% confidence level upper limit on the local burst rate density of \acp{PBH} \cite{Abdo:2014apa}:
\begin{equation}\label{eq:up99}
\mathrm{UL}_{99}=\frac{4.6}{V \times S}\ ,
\end{equation}  
which will be utilized in the studies presented in the next section. 
As noted, $V$ and $S$ denote the effective detectable volume and the observation duration, respectively.

\section{Prospective sensitivity of LHAASO to detect PBH bursts}\label{sec:3}

In this section, we present the expected constraints from \ac{LHAASO} on the burst rate density of \acp{PBH} in the vicinity of our Sun. {In addition, we propose data analysis strategies for LHAASO to optimize the search for PBHs and reach its potential detection limits.}

\begin{figure*}[htbp]
    \centering
    \includegraphics[width=0.6\textwidth]{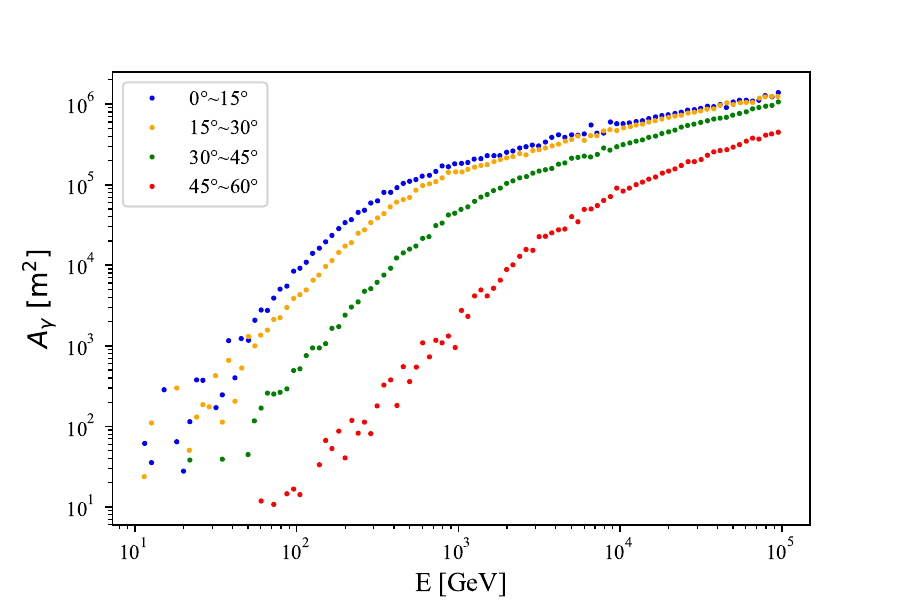}
    \includegraphics[width=0.6\textwidth]{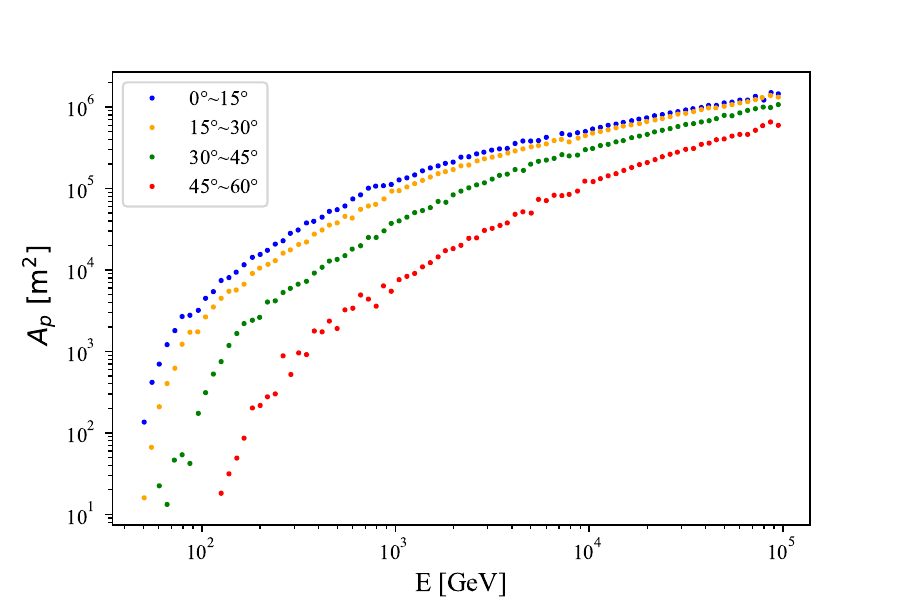}
    \includegraphics[width=0.6\textwidth]{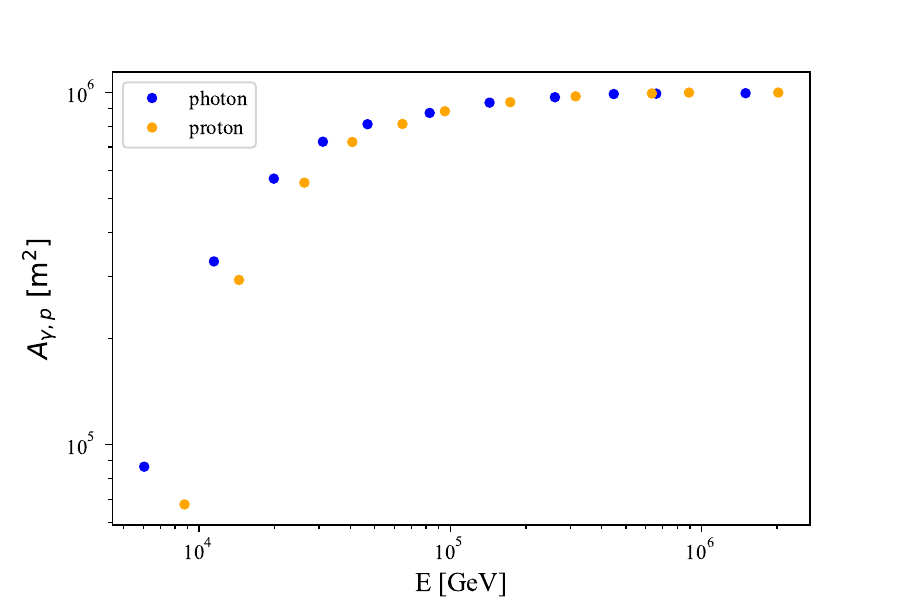}
    \caption{ Effective areas for photons and protons with respect to the energy bands and zenith angular bands. For WCDA (the zenith angle between $0^\circ$ and $60^\circ$ is equally divided into four bands), we show the upper panel for photons and the middle panel for protons. For KM2A (the zenith angle between $0^\circ$ and $45^\circ$ has only one band), we show the bottom panel for both photons and protons. These panels are reproduced from Refs.~\cite{Wang:2022jps,Cui:2014bda}. }
    \label{fig:figure1}
\end{figure*}

\begin{table*}[!h]
\centering

\setlength{\tabcolsep}{18pt}
\begin{tabularx}{\textwidth}{llllll}  
\hline
\hline
\text{Det.} & \textbf{$\tau\,[s]$} & \textbf{$\theta_i$} & \textbf{$N_{t}$} & \textbf{$n_\mathrm{bk}$} & \textbf{$\mu_{\min}$} \\

\midrule
\multirow{24}{*}{{WCDA}}
& $10^{-3}$ & $0^\circ - 15^\circ~(\theta_1)$ & $3.1 \times 10^{14}$ & 0.3 & 14 \\
& $10^{-3}$ & $15^\circ - 30^\circ~(\theta_2)$ & $3.1 \times 10^{14}$ & 0.2 & 12 \\
& $10^{-3}$ & $30^\circ - 45^\circ~(\theta_3)$ & $3.1 \times 10^{14}$ & 0.07 & 11 \\
& $10^{-3}$ & $45^\circ - 60^\circ~(\theta_4)$ & $3.1 \times 10^{14}$ & 0.01 & 8 \\
& $10^{-2}$ & $0^\circ - 15^\circ~(\theta_1)$ & $3.1 \times 10^{13}$ & 3 & 27\\  
& $10^{-2}$ & $15^\circ - 30^\circ~(\theta_2)$ & $3.1 \times 10^{13}$ & 2 & 24 \\ 
& $10^{-2}$ & $30^\circ - 45^\circ~(\theta_3)$ & $3.1 \times 10^{13}$ & 0.7 & 17\\ 
& $10^{-2}$ & $45^\circ - 60^\circ~(\theta_4)$ & $3.1 \times 10^{13}$ & 0.1 & 11 \\ 
& $10^{-1}$ & $0^\circ - 15^\circ~(\theta_1)$ & $3.1 \times 10^{12}$ & 30 & 60 \\ 
& $10^{-1}$ & $15^\circ - 30^\circ~(\theta_2)$ & $3.1 \times 10^{12}$ & 20 & 50\\ 
& $10^{-1}$ & $30^\circ - 45^\circ~(\theta_3)$ & $3.1 \times 10^{12}$ & 7 & 40 \\ 
& $10^{-1}$ & $45^\circ - 60^\circ~(\theta_4)$ & $3.1 \times 10^{12}$ & 1 & 20 \\ 
& 1 & $0^\circ - 15^\circ~(\theta_1)$ & $3.1 \times 10^{11}$ & 300& 170 \\  
& 1 & $15^\circ - 30^\circ~(\theta_2)$ & $3.1 \times 10^{11}$ & 200 & 140 \\  
& 1 & $30^\circ - 45^\circ~(\theta_3)$ & $3.1 \times 10^{11}$ & 70 & 80 \\  
& 1 & $45^\circ - 60^\circ~(\theta_4)$ & $3.1 \times 10^{11}$ & 10 & 40 \\  
& 10 & $0^\circ - 15^\circ~(\theta_1)$ & $3.1 \times 10^{10}$ & 3000 & 500 \\  
& 10 & $15^\circ - 30^\circ~(\theta_2)$ & $3.1 \times 10^{10}$ & 2000 & 400 \\  
& 10 & $30^\circ - 45^\circ~(\theta_3)$ & $3.1 \times 10^{10}$ & 700 & 200 \\  
& 10 & $45^\circ - 60^\circ~(\theta_4)$ & $3.1 \times 10^{10}$ & 100 & 100 \\  
& $10^{2}$ & $0^\circ - 15^\circ~(\theta_1)$ & $3.1 \times 10^{9}$ & 30000 & 1400 \\  
& $10^{2}$ & $15^\circ - 30^\circ~(\theta_2)$ & $3.1 \times 10^{9}$ & 20000 & 1100 \\
& $10^{2}$ & $30^\circ - 45^\circ~(\theta_3)$ & $3.1 \times 10^{9}$ & 7000 & 700 \\
& $10^{2}$ & $45^\circ - 60^\circ~(\theta_4)$ & $3.1 \times 10^{9}$ & 1000 & 300 \\
\midrule
\multirow{6}{*}{{KM2A}}
& $10^{-3}$ & $0^\circ - 45^\circ~(\theta_1)$ & $3.9 \times 10^{14}$ & 0.001 & 6 \\
& $10^{-2}$ & $0^\circ - 45^\circ~(\theta_1)$ & $3.9 \times 10^{13}$ & 0.01 & 7 \\  
& $10^{-1}$ & $0^\circ - 45^\circ~(\theta_1)$ & $3.9 \times 10^{12}$ & 0.1 & 11\\ 
& 1 & $0^\circ - 45^\circ~(\theta_1)$ &  $3.9 \times 10^{11}$ & 1 & 20\\  
& 10 & $0^\circ - 45^\circ~(\theta_1)$ &  $3.9 \times 10^{10}$ & 10 & 40 \\  
& $10^{2}$ & $0^\circ - 45^\circ~(\theta_1)$ &  $3.9 \times 10^{9}$ & 100 & 100 \\  

\hline
\hline
\end{tabularx}
\caption{\label{tab:tables1}The minimum number of counts $\mu_{\min}(\theta_i, \tau)$ required for a 5$\sigma$ detection with 50\% probability for a series of burst durations $\tau$ and zenith angle bands $\theta_{i}$ for WCDA and KM2A during a 3 year observing run. Here, we also show the number of trials $N_{t}$ and the number of background counts $n_{\mathrm{bk}}$.}
\end{table*}

The \ac{LHAASO} project consists of two detectors: the \ac{WCDA} and the \ac{KM2A} \cite{LHAASO:2019qtb}. 
The effective areas for photons and protons concerning the zenith angle bands for \ac{WCDA} can be found in Ref.~\cite{Wang:2022jps}, while those for \ac{KM2A} are provided in in Ref.~\cite{Cui:2014bda}. 
These effective areas are reproduced in Fig.~\ref{fig:figure1}. 
The field of view and the angular resolution of \ac{WCDA} are $\theta_{\mathrm{fov}}=120^\circ$ and $\theta_{\mathrm{res}}=2.1^\circ$, respectively, whereas for \ac{KM2A}, they are $\theta_{\mathrm{fov}}=90^\circ$ and $\theta_{\mathrm{res}}=1.4^\circ$. 
Based on these configurations and following the methodology outlined in Section~\ref{sec:2}, we obtain the explicit results for $N_{t}$, $n_{\mathrm{bk}}$, and $\mu_{\mathrm{min}}$ for each zenith angle band across several $\tau$'s by assuming a 3-year observing run. 
They are listed in Tab.~\ref{tab:tables1}.

\begin{table*}[!h]

\setlength{\tabcolsep}{35pt}
\begin{tabularx}{\textwidth}{llll}   
\hline
\hline
\text{Det.} & \textbf{$\tau\,[s]$} & \textbf{$r_{\max}\,[\mathrm{pc}]$} & \textbf{$\mathrm{UL}_{99}\,[\mathrm{pc}^{-3}\mathrm{yr}^{-1}]$}\\
\midrule
\multirow{6}{*}{{WCDA}} & $10^{-3}$ & 0.09 & 5100 \\
& $10^{-2}$ & 0.11  & 2300 \\  
& $10^{-1}$ & 0.12  & 1600\\ 
& 1 & 0.13  & 1600\\  
& 10 & 0.12  & 2000\\  
& $10^{2}$ & 0.11  & 3300\\  
\midrule
\multirow{6}{*}{{KM2A}}
& $10^{-3}$ & 0.08  & 5500\\ 
& $10^{-2}$ & 0.11  & 1800\\  
& $10^{-1}$ & 0.13  & 1200\\ 
& 1 & 0.12  & 1400\\  
& 10 & 0.09  & 4100\\  
& $10^{2}$ & 0.05  & 16000\\    
\hline
\hline
\end{tabularx}
\caption{\label{tab:table1} The 99\% confidence level upper limits on the local burst rate density of PBHs, i.e., $\mathrm{UL}_{99}$, for WCDA and KM2A for a 3 year observing run. For the burst durations in Tab.~\ref{tab:tables1}, we also show the maximum detectable distance $r_{\mathrm{max}}$, which corresponds to the zenith angle band $\theta_{1}$ in Tab.~\ref{tab:tables1}, and the effective detectable volume $V$.}

\end{table*}

\begin{figure*}[htbp]  
\centering\includegraphics[width=\textwidth]{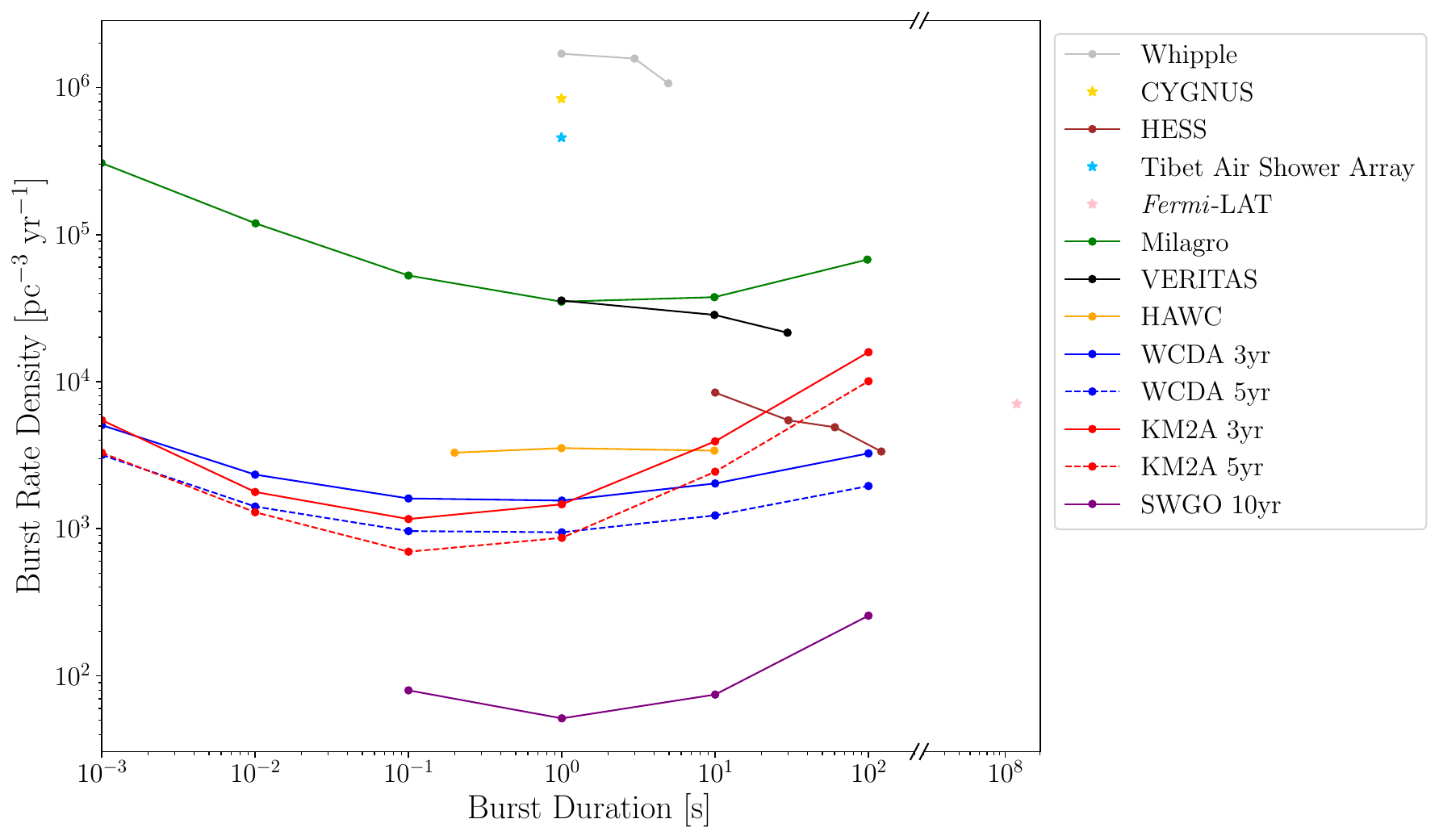} 
\caption{The 99\% confidence level upper limits on the local burst rate density of PBHs, i.e., $\mathrm{UL}_{99}$, for WCDA and KM2A for 3-year and 5-year observing runs. Observational constraints at 99\% confidence level from \ac{Whipple} \cite{Linton:2006yu}, \ac{CYGNUS} \cite{Alexandreas:1993zx}, \ac{H.E.S.S.} \cite{HESS:2023zzd}, Tibet Air Shower Array \cite{amenomori1995search}, \ac{Milagro} \cite{Abdo:2014apa}, \ac{VERITAS}  \cite{Archambault:2017asc}, {\ac{Fermi-LAT}} \cite{Fermi-LAT:2018pfs}, \acp{HAWC} \cite{HAWC:2019wla} and {\ac{SWGO}} \cite{Lopez-Coto:2021lxh} are shown for comparison.}
\label{fig:figure8}
\end{figure*}

Building upon the methodology outlined in Section~\ref{sec:2} and still assuming a 3-year observing run, we also derive explicit results for $r_{\mathrm{max}}$ (for $\theta_{1}$), and $\mathrm{UL}_{99}$ for the burst durations coinciding with those in Tab.~\ref{tab:tables1}. 
In addition, for 3-year and 5-year observing runs, the expected upper limits on the local burst rate density of \acp{PBH}, denoted as $\mathrm{UL}_{99}$, are depicted in Fig.~\ref{fig:figure8}. 
For comparison, we also depict the observational constraints from \ac{Whipple} \cite{Linton:2006yu}, \ac{CYGNUS} \cite{Alexandreas:1993zx}, \ac{H.E.S.S.} \cite{HESS:2023zzd}, Tibet Air Shower Array \cite{amenomori1995search}, \ac{Milagro} \cite{Abdo:2014apa}, \ac{VERITAS}  \cite{Archambault:2017asc}, {\ac{Fermi-LAT}}  \cite{Fermi-LAT:2018pfs}, and \acp{HAWC} \cite{HAWC:2019wla}, {as well as the predicted constraints from the planned detector \ac{SWGO} \cite{Lopez-Coto:2021lxh}.}
We find that \ac{LHAASO} exhibits superior sensitivity in detecting \ac{PBH} bursts compared to these existing programs.  
Following the 5-year observing run, it is expected to improve the upper limit reported by \ac{HAWC}, which is currently the best available, by an order of magnitude.

As illustrated in Fig.~\ref{fig:figure8}, the observing duration $S$ significantly affects our expected constraints on the local burst rate density of \acp{PBH}.
Although we focus on a 3-year observing duration in Tab.~\ref{tab:tables1}, $\mu_{\mathrm{min}}$ remains largely unchanged when considering other observing durations. 
In fact, it is only modified by $S$ in $N_{t}$. 
When examining other observing durations, such as 1 year or 10 years, we find that the post-trial probability in Eq.~(\ref{eq:posttrial}) changes only slightly. 
This results in only a minor adjustment to $\mu_{\mathrm{min}}$  \cite{Lopez-Coto:2021lxh}. 
Consequently, $r_{\mathrm{max}}$  remains constant, and thus $V$ does as well.  
Finally, while our results for $\mathrm{UL}_{99}$, shown in the last column of Tab.~\ref{tab:table1}, have been obtained using a 3-year observing run (i.e., blue solid curve for \ac{WCDA} and red solid curve for \ac{KM2A}), they can be readily adjusted for other durations by straightforwardly rescaling the value of $S$ in Eq.~(\ref{eq:up99}). 
Correspondingly, we can also modify the constraining curves in Fig.~\ref{fig:figure8} for a quick estimate. 
To validate this assertion, we further present in Fig.~\ref{fig:figure8} the exact results of $\mathrm{UL}_{99}$ for the 5-year observing run (i.e., blue dashed curve for \ac{WCDA} and red dashed curve for \ac{KM2A}), which coincide with the results from our quick estimate.

Finally, we emphasize that the {data analysis strategies} for the \ac{PBH} bursts is significantly influenced by the burst duration $\tau$.  
It is important to note that, compared with \ac{WCDA}, \ac{KM2A} is sensitive to a higher energy band \cite{Wang:2022jps,Cui:2014bda}, which corresponds to a higher temperature of \acp{PBH} and, consequently, a shorter burst duration. 
This explains why \ac{WCDA} is most sensitive to lasting $\sim1\,\mathrm{s}$ bursts while \ac{KM2A} is most sensitive to lasting $\sim0.1\,\mathrm{s}$ bursts, as demonstrated in Fig.~\ref{fig:figure8}. 
In this context, we suggest that \ac{WCDA} searches for \ac{PBH} bursts lasting  $\sim0.1-10\,\mathrm{s}$, while \ac{KM2A} searches for the \ac{PBH} bursts lasting  $\sim0.01-1\,\mathrm{s}$. 
We leave the study of realistic data analysis to future works. 

\section{Conclusions and discussion}\label{sec:5}

In this work, we investigated the capability of \ac{LHAASO} to search for the \ac{PBH} bursts in the vicinity of our Sun. 
We obtained the expected sensitivities of two detectors, \ac{WCDA} and \ac{KM2A}, to measure the local burst rate density of \acp{PBH}. 
Our findings revealed that both detectors can detect the \ac{PBH} bursts within a distance $\sim0.1\,\mathrm{pc}$ from the Sun. 
Conversely, the null detection would provide an upper limit on the local burst rate density of \acp{PBH}. 
For the 3-year (or 5-year) observing run of \ac{LHAASO}, we find the 99\% confidence level upper limit to be $\sim$ {1200} (or {700})$\,\mathrm{pc}^{-3}\mathrm{yr}^{-1}$, representing a sensitivity that is one order of magnitude stronger than the strongest observational constraint from \ac{HAWC}. 
{It is important to emphasize that the background rate $R_{b}(\theta_{i})$ calculated from Eq.~(\ref{eq:R}) is only a rough estimate, which may result in our findings being overly optimistic compared to the actual observed data. Observational results from \ac{HAWC} have confirmed this point \cite{Abdo:2014apa, HAWC:2019wla}.}
{Considering the duration of \ac{PBH} bursts, we propose data analysis strategies to optimize the search for these bursts and achieve the lowest possible limits.}
Specifically, we suggest that \ac{WCDA} and \ac{KM2A} target \ac{PBH} bursts with remaining lifetimes of $\sim0.1-10\,\mathrm{s}$ and $\sim0.01-1\,\mathrm{s}$, respectively.
Finally, we wish to emphasize the importance of searching for the \ac{PBH} bursts, as they are related not only to the quantum effects of gravity but also to fundamental cosmological questions, such as dark matter and the origin of the cosmos.

As a concluding remark, we compare \ac{LHAASO} with two future programs, namely the {\ac{CTA}} \cite{CTAConsortium:2017dvg} and \ac{SWGO} \cite{Albert:2019afb}. 
Although limited by its field of view, {\ac{CTA}} can reach a larger $r_{\mathrm{max}}$ due to its superior sensitivity and is therefore expected to remain competitive with \ac{LHAASO}. 
With its wide field of view, \ac{SWGO} is projected to measure the local burst rate density at approximately 50 $\mathrm{pc}^{-3}\mathrm{yr}^{-1}$ during a ten-year observing run \cite{Lopez-Coto:2021lxh}, as shown in Fig~\ref{fig:figure8}.
This sensitivity makes it competitive with that of \ac{LHAASO}.
However, we emphasize that \ac{LHAASO} monitors the northern hemisphere while \ac{SWGO} monitors the southern hemisphere. 
They are complementary rather than competitive.

\acresetall


\vspace{1em}
\acknowledgments

We thank Xiao-Jun Bi, Peng-Fei Yin, Shou-Shan Zhang, Xu-Kun Zhang, and Zhi-Chao Zhao for helpful discussions. 
This work is supported by the National SKA Program of China (Grant Nos. 2022SKA0110200 and 2022SKA0110203), the National Natural Science Foundation of China (Grant Nos. 12473001, 12175243, and 11975072), the National Key R\&D Program of China (Grant No. 2023YFC2206403), the Science Research Grants from the China Manned Space Project (Grant No. CMS-CSST-2021-B01), the Key Research Program of the Chinese Academy of Sciences (Grant No. XDPB15), and the 111 Project (Grant No. B16009).

\bibliographystyle{JHEP}
\bibliography{lhaaso}

\end{document}